\documentclass[12pt,a4paper]{article}
\usepackage{a4wide}
\usepackage{epsfig}
\usepackage{amsmath}
\usepackage{graphicx}
\usepackage{latexsym}
\usepackage{array}

  \newlength{\absize}
\newcommand{\Lumint}{{\cal L}_{\rm int}}

\begin{document}

  \thispagestyle{empty}
  \renewcommand{\thefootnote}{\fnsymbol{footnote}}
\newpage\normalsize
    \pagestyle{plain}
    \setlength{\baselineskip}{4ex}\par
    \setcounter{footnote}{0}
    \renewcommand{\thefootnote}{\arabic{footnote}}
\newcommand{\preprint}[1]{%
  \begin{flushright}
    \setlength{\baselineskip}{3ex} #1
  \end{flushright}}
\renewcommand{\title}[1]{%
  \begin{center}
    \LARGE #1
  \end{center}\par}
\renewcommand{\author}[1]{%
  \vspace{2ex}
  {\Large
   \begin{center}
     \setlength{\baselineskip}{3ex} #1 \par
   \end{center}}}
\renewcommand{\thanks}[1]{\footnote{#1}}
\renewcommand{\abstract}[1]{%
  \vspace{2ex}
  \normalsize
  \begin{center}
    \centerline{\bf Abstract}\par
    \vspace{2ex}
    \parbox{\absize}{#1\setlength{\baselineskip}{2.5ex}\par}
  \end{center}}

\vspace*{4mm} 

\title{Distinguishing new physics scenarios with \\  polarized electron and
positron beams} \vfill

\author{A. A. Pankov,$^{a,}$\footnote{E-mail: pankov@ictp.it} N. Paver$^{b}$ and A. V. Tsytrinov$^{a}$}

\begin{center}
$^{a}$ICTP Affiliated Centre, Pavel Sukhoi Technical University,
Gomel 246746, Belarus\\
$^{b}$University of Trieste and INFN-Trieste Section, 34100
Trieste, Italy
\end{center}

%
%
%
\vfill \abstract{Contactlike nonstandard interactions can be
revealed only through deviations of observables from the standard
model (SM) predictions. We consider a number of such nonstandard
scenarios, and discuss their identification as sources of
deviations in fermion-pair production processes at the
International Linear Collider (ILC), if they were observed. We
emphasize the r\^ole of $e^-$ {\it and} $e^+$ polarization in
enhancing the identification reaches.}

\vspace*{20mm} \setcounter{footnote}{0} \vfill

\begin{center}
Talk given at the \textit{International Linear Collider Workshop -
LCWS06},\\ Bangalore, India, 9-13 March, 2006.
\end{center}

\newpage
    \setcounter{footnote}{0}
    \renewcommand{\thefootnote}{\arabic{footnote}}

\section{Introduction}
New physics (NP) beyond the standard model is expected to show up at the LHC
and ILC colliders either {\it directly} through production of new particles,
or {\it indirectly} through {\it deviations} of cross sections and
asymmetries from the SM predictions. The latter case is typical of
interactions mediated by exchanges of heavy mass objects, such that the
energy is not sufficient for their direct production. The corresponding
corrections to the SM predictions are most conveniently parameterized in
terms of negative powers of the characteristic large mass
scales $\Lambda$, times matrix elements of effective, contactlike, Lagrangians.

The so-called {\it discovery reach}, i.e., the maximum value of $\Lambda$ for
which a deviation could be observed within the foreseeable experimental
accuracy, gives an indication of the expected sensitivity of an observable
to the various NP scenarios. On the other hand, in principle different NP
models can cause similar deviations. Therefore, it should be interesting to
assess the {\it identification reach} on the individual models, i.e., the
maximum value of $\Lambda$ for which a novel interaction not only produces
observable deviations, but also can be discriminated, as the source of the
observed deviations, from the other nonstandard interactions for all values
of their characteristic mass scale parameters. Clearly, by definition, the
identification reach is expected to be smaller than the discovery reach.
Here we discuss the differential cross sections for the following processes
at the ILC with both beams longitudinally polarized:
\begin{equation}
e^++e^-\to{\bar f}+f,\qquad\quad f=e,\mu,\tau,c,b\, .
\label{proc}
\end{equation}
These processes can all receive corrections from the contactlike interactions
considered here, and their sensitivity is significantly enhanced by the
initial $e^-$ and $e^+$ polarizations
\cite{Pankov:2005kd,Pankov:2005ar,Osland:2003fn,Pankov:2002qk}. This facility
is envisaged at the planned ILC \cite{Moortgat-Pick:2005cw}.


The nonstandard scenarios we consider are the following:

{\it a)} The ADD large compactified extra dimensions scenario
\cite{Arkani-Hamed:1998rs,Arkani-Hamed:1998nn,Antoniadis:1998ig}, where
only gravity can propagate in the extra spatial dimensions, and
correspondingly a tower of graviton KK states is exchanged in the
four-dimensional space \cite{Giudice:1998ck,Han:1998sg}. The
relevant, dimension-8, effective Lagrangian can be expressed
as \cite{Hewett:1998sn}:
\begin{equation}
{\cal L}^{\rm
ADD}=i\frac{4\lambda}{\Lambda_H^4}T^{\mu\nu}T_{\mu\nu},
\label{dim-8}
\end{equation}
with $T_{\mu\nu}$ being the energy-momentum tensor of the SM particles and
$\Lambda_H$ a phenomenological cutoff on the summation over the KK spectrum,
expected in the (multi) TeV region. Here, $\lambda=\pm1$.

{\it b)} Gravity in ${\rm TeV}^{-1}$--scale extra dimensions, in
which also the SM gauge bosons can propagate. The relevant
contactlike effective Lagrangian can be parameterized by a
``compactification scale'' $M_C$, and for one extra dimension it
reads \cite{Cheung:2001mq,Rizzo:1999br}:
\begin{eqnarray}
{\cal L}^{\rm TeV}&=&-\frac{\pi^2}{3M_C^2}
[Q_e Q_f({\bar e}\gamma_\mu e)({\bar f}\gamma^\mu f)
\nonumber \\
&+&(g_{\rm L}^e{\bar e}_{\rm L}\gamma_\mu e_{\rm L}
+g_{\rm R}^e{\bar e}_{\rm R}\gamma_\mu e_{\rm R})
(g_{\rm L}^f{\bar f}_{\rm L}\gamma^\mu f_{\rm L}
+g_{\rm R}^f{\bar f}_{\rm R}\gamma^\mu f_{\rm R})].
\label{tev}
\end{eqnarray}

{\it c)} The dimension-6 four-fermion contact interactions (CI)
\cite{Eichten:1983hw,Ruckl:1983hz}, with
$\Lambda_{\alpha\beta}$ ``compositeness'' mass scales
($\alpha,\beta={\rm L,R}$ and $\eta_{\alpha\beta}=\pm1,0$):
\begin{equation}
{\cal L}^{\rm CI}=\frac{4\pi}{1+\delta_{ef}}\hskip 3pt \sum_{\alpha,\beta}
\hskip 3pt
\frac{\eta_{\alpha\beta}}{\Lambda^2_{\alpha\beta}} \left(\bar
e_\alpha\gamma_\mu e_\alpha\right) \left(\bar f_\beta\gamma^\mu
f_\beta\right).
\label{CI}
\end{equation}

Current experimental lower bounds on the above mass scales at the 95\% C.L.
are $\Lambda_H>1.1-1.3\, {\rm TeV}$ and $M_C>6.8\, {\rm TeV}$
\cite{Cheung:2004ab} while, generically, those on $\Lambda$s of
Eq.~(\ref{CI}) are in the range 10--15 TeV \cite{Eidelman}.
\section{Derivation of discovery reaches}


\begin{table}[!htb]

\caption{95\% C.L. discovery reaches (in TeV). Left and right
entries refer to the polarization configurations $(\vert
P^-\vert,\vert P^+\vert$)=(0,0) and (0.8,0.6), respectively.
\label{table:DIS}}
\begin{center}
\begin{tabular}{lcccc} \hline
\raisebox{-1.50ex}[0cm][0cm]{Model}&
\multicolumn{4}{c}{Process}  \\
 &
$e^{+}e^{-} \to e^{+}e^{-}$ & $e^{+}e^{-} \to l^{+}l^{-}$
& $e^{+}e^{-} \to \bar{b}b$ & $e^{+}e^{-} \to \bar{c}c$  \\ \hline \\[-0.3cm]
 & \multicolumn{4}{c}{ $\sqrt{s} = 0.5$ TeV, $\Lumint = 100fb^{-1}$}  \\
\cline{2-5}
$ \Lambda _{H} $ & 4.1; 4.3& 3.0; 3.2& 3.0; 3.4& 3.0; 3.2 \\
$\Lambda _{VV}^{ef}$& 76.2; 86.4& 89.7; 99.4& 76.1; 96.4& 84.0; 94.1 \\
$\Lambda _{AA}^{ef}$& 47.4; 69.1& 80.1; 88.9& 76.7; 98.2& 76.5; 85.9 \\
$\Lambda _{LL}^{ef}$& 37.3; 52.5& 53.4; 68.3& 63.6; 72.7& 54.5; 66.1 \\
$\Lambda _{RR}^{ef}$& 36.0; 52.2& 51.3; 68.3& 42.5; 71.2& 46.3; 66.8 \\
$\Lambda _{LR}^{ef}$& 59.3; 69.1& 48.5; 62.8& 51.3; 68.7& 37.0; 57.7 \\
$\Lambda _{RL}^{ef}$& $\Lambda _{RL}^{ee} = \Lambda _{LR}^{ee}$
& 48.7; 63.6& 46.8; 60.1& 52.2; 60.7 \\
$M_{C}$& 12.0; 13.8& 20.0; 22.2& 6.6; 10.7& 10.4; 12.0

\end{tabular}
\end{center}
\end{table}

For an extensive presentation of the analysis and a full account
of the numerical results for the expected discovery and
identification reaches we refer to \cite{Pankov:2005kd}.
Basically, for the polarized angular distributions,
${\cal O}={\rm d}\sigma/{\rm d}\cos\theta$, we introduce the relative
deviations from the SM predictions and the corresponding $\chi^2$:
\begin{equation}
\Delta ({\cal O})=\frac{{\cal O}(\rm SM+NP)-{\cal O}(\rm SM)}{{\cal O}
(\rm SM)};\qquad
\chi^2({\cal O})= \sum_{\rm bins}\left(\frac{\Delta({\cal O})^{\rm
bin}} {\delta{\cal O}^{\rm bin}}\right)^2.
\label{reldev}
\end{equation}
Here, the angular range has been divided into ten equal-size bins,
and $\delta{\cal O}^{\rm bin}$ denotes the expected relative
uncertainty, statistical plus systematic ones, in each bin. The
discovery reaches on models (\ref{dim-8})-(\ref{CI}) can be
assessed by assuming nonobservation of deviations and,
accordingly, are determined by the condition $\chi^2({\cal
O})<\chi^2_{\rm CL}$. Here, we take $\chi^2_{\rm CL}=3.84$ for a
95\% C.L. In Table~\ref{table:DIS} we present numerical results
for an ILC with parameters as specified in the caption. The
assumed reconstruction efficiencies are 100\% for $e^+e^-$ final
pairs; 95\% for $l^+l^-$ events ($l=\mu,\tau$); 35\% and 60\% for
$c {\bar c}$ and $b {\bar b}$, respectively. The major systematic
uncertainties originate from uncertainties on beams polarizations
and on the time-integrated luminosity, for which we have assumed
$\delta P^-/P^-=\delta P^+/P^+=0.2$\% and $\delta{\cal L}_{\rm
int}/{\cal L}_{\rm int}=0.5$\%, respectively. The results in
Table~\ref{table:DIS} clearly show the enhancement in sensitivity
to NP models allowed, at a given C.M. energy, by beams
polarization. In particular, this effect is dramatic in the case
of the CI models (\ref{CI}).

\section{Assessment of expected identification reaches}
In Ref.~{\cite{Pasztor:2001hc}, the identification reaches at ILC on
contactlike interactions were obtained from a Monte Carlo-based analysis of the
unpolarized leptonic processes (\ref{proc}). Here we consider polarized
beams and also quark final states. Continuing the previous $\chi^2$-based
analysis, we now assume that a deviation has been observed, for example
consistent with the ADD scenario (\ref{dim-8}) with some value of
$\Lambda_H$. To assess the level at which the ADD model can be identified
from the other models potential sources of the deviation, we choose, for
example, the AA model of Eq.~(\ref{CI}), and introduce relative deviations
${\tilde\Delta}$ and corresponding ${\tilde\chi}^2$ similar to
Eq.~(\ref{reldev}):
\begin{equation}
{\tilde\Delta} ({\cal O})= \frac{{\cal O}({\rm AA})-{\cal O}({\rm
ADD})}{{\cal O}({\rm ADD})}; \qquad\quad {\tilde\chi}^2({\cal O})=
\sum_{\rm bins}\left (\frac{{\tilde\Delta}({\cal O})^{\rm bin}}
{{\tilde\delta}{\cal O}^{\rm bin}}\right)^2. \label{chitilde}
\end{equation}


\begin{table}[!htb]
 \caption{95\% C.L. identification reaches (in
TeV). Left and right entries refer to the polarization
configurations ($|P^+|$,$|P^-|$)=(0,0) and (0.8, 0.6).
\label{table:IDR}}
\begin{center}
\begin{tabular}{lcccc}
\hline \raisebox{-1.50ex}[0cm][0cm]{Model}& \multicolumn{4}{c}{Process}  \\
 & $e^{+}e^{-} \to e^{+}e^{-}$ & $e^{+}e^{-} \to l^{+}l^{-} $
& $e^{+}e^{-} \to \bar{b}b$ & $e^{+}e^{-} \to \bar{c}c$ \\
\hline \\[-0.3cm] &
\multicolumn{4}{c}{ $\sqrt{s} = 0.5$ TeV, $\Lumint = 100fb^{-1}$}
\\ \cline{2-5}
$ \Lambda _{H} $ & 2.5; 3.1& 2.7; 2.8& 3.0; 3.1& 2.7; 2.9 \\
$\Lambda _{VV}^{ef}$ & 26.8; 41.0& 57.6; 63.9& 20.3; 82.7& 56.3; 64.4 \\
$\Lambda _{AA}^{ef}$ & 41.1; 60.9& 63.2; 70.2& 20.4; 84.7& 62.2; 75.5 \\
$\Lambda _{LL}^{ef}$ & \hskip 5pt --- ; 44.3& \hskip 5pt --- ; 56.0& \hskip 5pt --- ; 62.2& \hskip 5pt --- ; 59.0 \\
$\Lambda _{RR}^{ef}$ & \hskip 5pt --- ; 45.2& \hskip 5pt --- ; 58.0& \hskip 5pt --- ; 61.1& \hskip 5pt --- ; 59.9 \\
$\Lambda _{LR}^{ef}$ & 31.3; 48.5& \hskip 5pt --- ; 57.5& \hskip 5pt --- ; 49.0& \hskip 5pt --- ; 46.9 \\
$\Lambda _{RL}^{ef}$ & $\Lambda _{RL}^{ee} = \Lambda _{LR}^{ee}$ & \hskip 5pt --- ; 58.3& \hskip 5pt --- ; 56.2& \hskip 5pt --- ; 53.1 \\
$ M_{C} $ & 4.2; 6.5& \hskip 6pt 8.8; 14.4& 4.5; 7.7& 6.0; 8.3

\end{tabular}
\end{center}
\end{table}
In Eq.~(\ref{chitilde}), ${\tilde\delta{\cal O}^{\rm bin}}$ is the
expected relative uncertainty, the statistical part being related
to the ADD model prediction. Accordingly, ${\tilde\chi}^2$ is a
function of $\lambda/\Lambda_H^4$ and $\eta/\Lambda_{\rm AA}^2$,
and we can determine, in the plane of these parameters, the {\it
confusion region} where the AA model can be considered as
consistent with the ADD. At the 95\%~C.L., one determines such
confusion region from the condition ${\tilde\chi}^2<3.84$. The
contour of the confusion region identifies a maximal value of
$\vert\lambda/\Lambda_H^4\vert$ (equivalently, a minimum value of
$\Lambda_H$), for which the AA  model can {\it excluded} at the 95
\% C.L. for any value of $\eta/\Lambda_{\rm AA}^2$. This value,
$\Lambda_H^{\rm AA}$, is the {\it exclusion reach} on the AA
model. Fig.~\ref{fig:1} (left panel) shows an example of confusion
region obtained by different polarization. This procedure can be
repeated for all other interactions in Eqs.~(\ref{tev}) and
(\ref{CI}), to determine the individual exclusion reaches
$\Lambda_H^{\rm VV}$, $\Lambda_H^{\rm RR}$, $\Lambda_H^{\rm LL}$,
$\Lambda_H^{\rm LR}$, $\Lambda_H^{\rm RL}$ and $\Lambda_H^{\rm
TeV}$. Finally, the {\it identification reach} on the ADD scenario
corresponds to the {\it minimum} of the {\it exclusion reaches},
$\Lambda_H^{\rm ID}=min\{\Lambda_H^{VV},\, \Lambda_H^{AA},
\Lambda_H^{RR},\, \Lambda_H^{LL},\, \Lambda_H^{LR},\,
\Lambda_H^{RL},\, \Lambda_H^{TeV}\}$. Clearly, for
$\Lambda_H<\Lambda_H^{\rm ID}$ all composite-like CI models as
well as the ${\rm TeV}^{-1}$ gravity model can be excluded as
explanations of the deviation or, equivalently, the ADD model can
be identified. This simple $\chi^2$ procedure can be applied in
turn to all the individual sources of corrections to the SM in
Eqs.~(\ref{tev}) and (\ref{CI}), and the expected identification
reaches on the corresponding $\Lambda$ mass parameters can be
determined analogously. Fig.~\ref{fig:1} (right panel) shows an
example, relevant to CI models, where a restricted confusion
region can be determined only with initial beams polarization.

\begin{figure}[!htb] 
\vspace*{0.0cm} \centerline{ \hspace*{-0.2cm}
\includegraphics[width=6.0cm,angle=0]{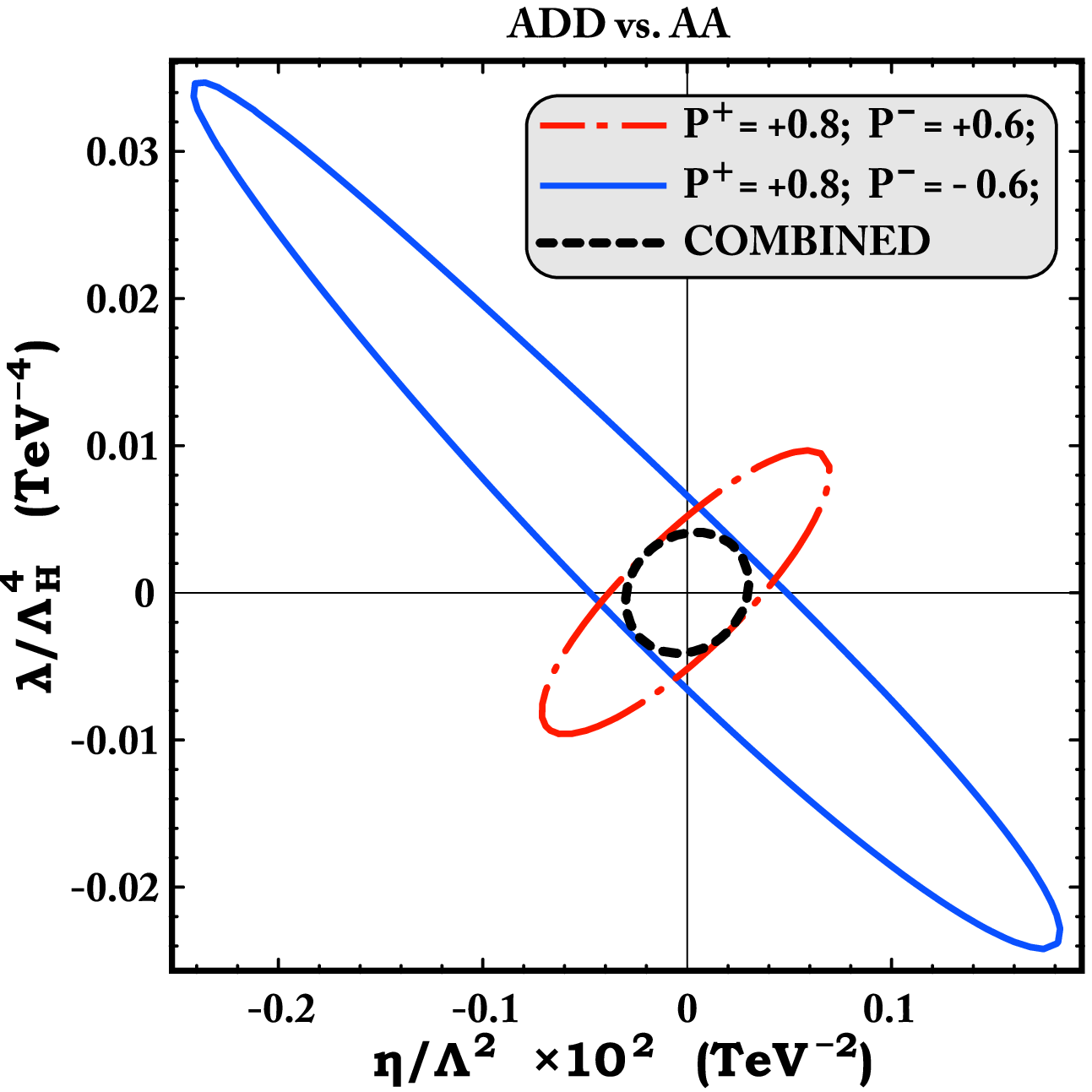}
\hspace*{0.2cm}
\includegraphics[width=6.0cm,angle=0]{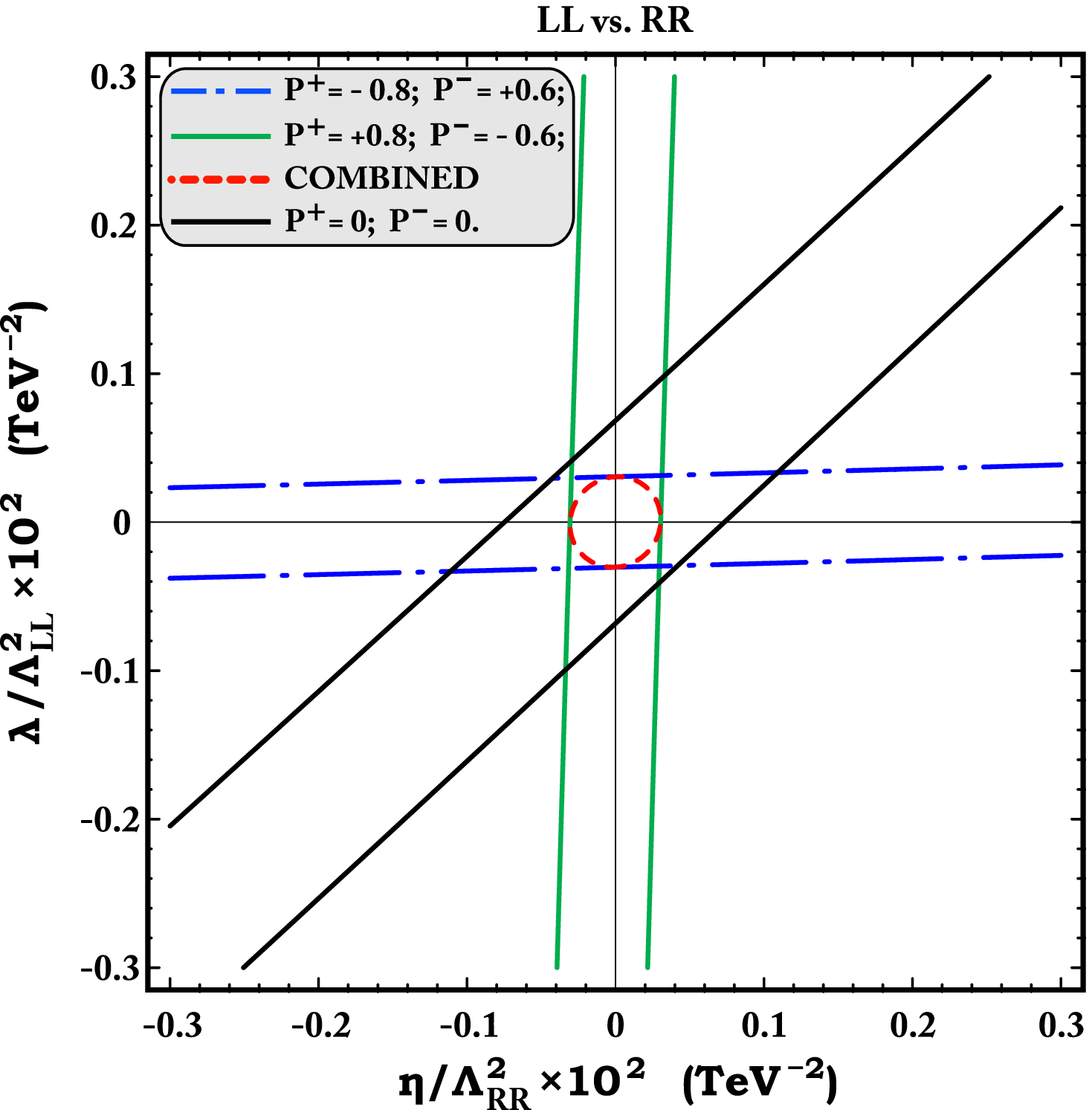}}
\vspace*{0.1cm}
\caption{\label{fig:1} Left panel: $\Lambda_{\rm H}$ {\it vs.}
$\Lambda_{\rm AA}$ confusion region from $e^+e^-\to e^+e^-$.
Right panel: $\Lambda_{\rm LL}$ {\it vs.} $\Lambda_{\rm RR}$ confusion region
from $e^+e^-\to l^+l^-$ ($l=\mu,\tau$).
$\sqrt{s}= 0.5 {\rm TeV}, {\cal L}_{\rm int}=100\, {\rm fb}^{-1}$. }
\end{figure}

The numerical results for the expected identification reaches are
shown in Table~\ref{table:IDR}. Here, blank entries refer to
models for which the identification reach is found to fall below
the current limits.
\section{Concluding remarks}
In the previous section we have considered the problem of
distinguishing the New Physics scenarios represented by the
contactlike effective Lagrangians (\ref{dim-8})-(\ref{CI}) from
one another at the ILC, by analyzing polarized differential cross
sections for fermion-pair production processes. The discovery
reaches as well as the identification reaches are rather high
compared to the current bounds, depending on energy and
luminosity, and can increase by factors 2.5-3 for an ILC with
$\sqrt s=1\, {\rm TeV}$ and $\Lumint = 1000\, {\rm fb}^{-1}$
\cite{Pankov:2005kd}. The r\^ole of
polarization in the various cases is shown by Fig.~\ref{fig:1} and
Table~\ref{table:IDR}. It has an appreciable r\^ole in enhancing
the identification sensitivity to the ADD model
(Fig.~\ref{fig:1}, left panel). The enhancement
is dramatic for models (\ref{tev}) and (\ref{CI}). Actually, as
indicated  by the blank entries in Table~\ref{table:IDR},
polarization should be essential for the identification of some of
the contact interaction scenarios (\ref{CI}), which would not be
possible with unpolarized beams (Fig.~\ref{fig:1}, right panel).



\begin{thebibliography}{99}

\bibitem{Pankov:2005kd}
A.~A.~Pankov, N.~Paver and A.~V.~Tsytrinov,
Phys.\ Rev.\ D {\bf 73}, 115005 (2006).

\bibitem{Pankov:2005ar}
  A.~A.~Pankov and N.~Paver,
  Phys.\ Rev.\ D {\bf 72}, 035012 (2005).

\bibitem{Osland:2003fn}
P.~Osland, A.~A.~Pankov and N.~Paver,
Phys.\ Rev.\ D {\bf 68}, 015007 (2003).

\bibitem{Pankov:2002qk}
 A.~A.~Pankov and N.~Paver,
 Eur.\ Phys.\ J.\ C {\bf 29}, 313 (2003).

\bibitem{Moortgat-Pick:2005cw}
G.~Moortgat-Pick {\it et al.},
arXiv:hep-ph/0507011.

\bibitem{Arkani-Hamed:1998rs}
N.~Arkani-Hamed, S.~Dimopoulos and G.~R.~Dvali,
Phys.\ Lett.\ B {\bf 429}, 263 (1998).

\bibitem{Arkani-Hamed:1998nn}
N.~Arkani-Hamed, S.~Dimopoulos and G.~R.~Dvali,
Phys.\ Rev.\ D {\bf 59}, 086004 (1999).

\bibitem{Antoniadis:1998ig}
I.~Antoniadis, N.~Arkani-Hamed, S.~Dimopoulos and G.~R.~Dvali,
Phys.\ Lett.\ B {\bf 436}, 257 (1998).

\bibitem{Giudice:1998ck}
G.~F.~Giudice, R.~Rattazzi and J.~D.~Wells,
Nucl.\ Phys.\ B {\bf 544}, 3 (1999).

\bibitem{Han:1998sg}
T.~Han, J.~D.~Lykken and R.~J.~Zhang,
Phys.\ Rev.\ D {\bf 59}, 105006 (1999).


\bibitem{Hewett:1998sn}
J.~L.~Hewett,
Phys.\ Rev.\ Lett.\  {\bf 82}, 4765 (1999).

\bibitem{Cheung:2001mq}
K.~M.~Cheung and G.~Landsberg,
Phys.\ Rev.\ D {\bf 65}, 076003 (2002).

\bibitem{Rizzo:1999br}
T.~G.~Rizzo and J.~D.~Wells,
Phys.\ Rev.\ D {\bf 61}, 016007 (2000).

\bibitem{Eichten:1983hw}
E.~Eichten, K.~D.~Lane and M.~E.~Peskin,
Phys.\ Rev.\ Lett.\  {\bf 50}, 811 (1983).

\bibitem{Ruckl:1983hz}
R.~R\"uckl,
Phys.\ Lett.\ B {\bf 129}, 363 (1983).

\bibitem{Cheung:2004ab}
For a review see, e.g., K.~Cheung,
arXiv:hep-ph/0409028.

\bibitem{Eidelman}
S.~Eidelman {\it et al.} [Particle Data Group],
Phys.\ Lett.\ B {\bf 502}, 1 (2004).

\bibitem{Pasztor:2001hc}
G.~Pasztor and M.~Perelstein,
arXiv:hep-ph/0111471.


\end{thebibliography}
\end{document}